\documentclass{libtex/llncs}
\usepackage{epsfig}
\usepackage{amsfonts}
\usepackage{libtex/srcltx} 
\usepackage[greek, american]{babel} 
\usepackage[iso-8859-7]{inputenc}

\long\def\comment#1{}
\long\def\comment#1{}

\newlength{\figtblfootnotemargin}
\newlength{\figtblfootnotewidth}

\newcommand{\outcomment}[1]{
    \comment{
        #1
    }
}

\newcommand{\SquareOfSize}[2]{
 \fbox{\hsize #1cm \hbox to #1cm{\vbox{#2}}}
}

\newcommand{\putsfigure}[2]{
        \psfig{figure=#1,height=#2}
}

\newcommand{\putfigureSV}[5]{

    \begin{figure}[htbp]
        \vspace*{-0.1cm}
        \centerline{ \psfig{figure=#1,height=#2}}
        \vspace*{-0.3cm}
        \caption{#4}{\vspace*{0.02cm}  \centerline{ \parbox[t]{10cm}{ #5}}}
        \label{#3}
        \vspace*{-0.4cm}
    \end{figure}
}

\newcommand{\putTwoFiguresEngl}[6]{
    \begin{figure}
        \begin{minipage}[b]{6cm}
            \centerline{\putsfigure{#1}{#3}}
            \vspace*{3mm} \centerline{(a)}
        \end{minipage}
        \begin{minipage}[b]{6cm}
            \centerline{\putsfigure{#2}{#3}}
            \vspace*{3mm} \centerline{(b)}
        \end{minipage}
        \caption{#5}{\vspace*{0.1cm}
            \centerline{ \parbox[t]{13cm}{\footnotesize #6}}}
        \label{#4}
        \vspace*{-5mm} 
    \end{figure}
}

\newcommand{\puttableSV}[4]{  
    \begin{table}[htbp]
        \begin{center}
            \vspace*{-1mm} 
            \caption{#2}{\vspace*{-3mm} \centerline{ \parbox[t]{10cm}{ #3}}}
            \vspace*{-3mm} 
            \label{#1}
            #4
        \end{center}
    \end{table}
    \vspace*{-2mm} 
}

\newsavebox{\wholeWidthLine}
\sbox{\wholeWidthLine} {\rule[0.1in]{\textwidth}{.01in}}

\newcommand{\bq}{\begin{quote}}
\newcommand{\eq}{\end{quote}}
\newcommand{\be}{\begin{enumerate}}
\newcommand{\ee}{\end{enumerate}}
\newcommand{\bi}{\begin{itemize}}
\newcommand{\ei}{\end{itemize}}
\newcommand{\bie}{\begin{itemize}\begin{enumerate}}
\newcommand{\eie}{\end{enumerate}\end{itemize}}
\newcommand{\ba}{\begin{array}}
\newcommand{\ea}{\end{array}}
\newcommand{\btbl}{\begin{tabular}}
\newcommand{\etbl}{\end{tabular}}
\newcommand{\bequ}{\begin{displaymath}}
\newcommand{\eequ}{\end{displaymath}}
\newcommand{\bequa}{\begin{eqnarray*}}
\newcommand{\eequa}{\end{eqnarray*}}
\newcommand{\bc}{\begin{center}}
\newcommand{\ec}{\end{center}}
\newcommand{\btab}{\begin{tabbing}}
\newcommand{\etab}{\end{tabbing}}

\def\mpr#1{\ifmmode #1 \else #1 \fi}

\newcommand{\DPATTRNAME}{{\large{{\bf SC}$_{attr}$}}}
\newcommand{\DPISANAME}{{\large{{\bf SC}$_{isA}$}}}
\newcommand{\DPINNAME}{{\large{{\bf SC}$_{in}$}}}

\include{paper}

\newcommand{\groogle}{\textit{\textsf{Mitos}}}

\begin{document}

\title{The Anatomy of \groogle\ Web Search Engine}
        \author{Panagiotis Papadakos, Giorgos Vasiliadis, Yannis Theoharis,\\
		Nikos Armenatzoglou, Stella Kopidaki, Yannis Marketakis,\\
		Manos Daskalakis, Kostas Karamaroudis, Giorgos Linardakis,\\
		Giannis Makrydakis, Vangelis Papathanasiou, Lefteris Sardis,\\
		Petros Tsialiamanis, Georgia Troullinou, Kostas Vandikas,\\
		Dimitris Velegrakis and Yannis Tzitzikas}
        \institute{
            Computer Science Department, University of Crete, GREECE, and \\
            Institute of Computer Science, FORTH-ICS, GREECE \\
            \texttt{\small\{papadako,gvasil,theohari,tzitzik\}@ics.forth.gr}
        \texttt{\small\{armenan,skopidak,marketak,mdaskal,karamar,linard,makrydak,epapath,\\
                        sardis,troulin,tsialiam,vandikas,velegrak\}@csd.uoc.gr}
}

\maketitle

\begin{abstract}
    Engineering a Web search engine offering
    effective and efficient information retrieval is a challenging task.
    This document presents our experiences from designing and
    developing a Web search engine offering a wide spectrum of functionalities
    and we report some interesting experimental results.
    A rather peculiar design choice of the engine is that its index is based on
    a DBMS,
    while some of the distinctive functionalities that are offered include
    advanced Greek language stemming,
    real time result clustering,
    and advanced link analysis techniques (also for spam page detection).

\end{abstract}

\section{Introduction}

Engineering a Web search engine offering effective and efficient 
information retrieval is a challenging task.
The objective of this document is to present our experiences from designing and
developing \groogle\ (previously named groogle),
a currently emerged search engine
that offers  a wide spectrum of functionalities.
Distinctive functionalities include:
(a) an advanced stemmer for the Greek language,
(b) real time result clustering,
(c) advanced link analysis techniques (also for spam detection).

Apart from describing the overall architecture and component
specifications, we report some interesting experimental results
regarding all associated tasks (i.e. crawling, stemming, stopwords
elimination,  indexing, query evaluation, clustering).
One crucial design choice of \groogle\ is that its index is based on a  DBMS.
This choice has some benefits and some drawbacks.
An advantage is that we can exploit all amenities offered by a DBMS.
To the problem at hand,  some benefits include:
(a) efficient index update
(e.g. efficient deletion of those index entries that concern web pages
that no longer exist, or addition of newly crawled or submitted web pages),
and
(b) straightforward extension of the index schema with additional columns and relations
 (e.g. for widening  the spectrum of functionalities offered).
The drawbacks include higher storage space requirements
and poorer performance in certain tasks.
To clarify this aspect, we compare our engine with other well-known inverted
file-based IR systems (like Terrier) and discuss the results of this
comparison.

The rest of this  paper is organized as follows:
Section \ref{sec:Arch} describes the overall architecture of the engine.
Section \ref{sec:Components} describes briefly each component.
Section \ref{sec:RW} reports experimental results,
and finally,
Section \ref{sec:Concl} concludes the paper and
identifies issues for further work and research.

\outcomment{
    This document has  an educational value
because there are only few
concise papers (e.g. \cite{arasu2001sw})
that discuss all aspects
revolving the design and implementation of Web search engines.
}
\section{Architectural Overview}
\label{sec:Arch}

This section gives a high level overview of
the architecture and the functioning of the system.
Most components of \groogle\  are implemented in Java 5
to preserve platform-independent deployment.
The current installation of \groogle\ runs on a single machine
(Pentium IV 3.2GHz, 2GB RAM, Debian) and is
accessible through http://groogle.csd.uoc.gr:8080/mitos/.

Two are the basic processes:
the indexing process that is performed offline and the
searching process that is performed every time a user
is searching for something based on a query.
Figure \ref{fig:groogleArchitecture} illustrates
the component diagram
and outlines the
indexing and retrieval process.

\putfigureSV{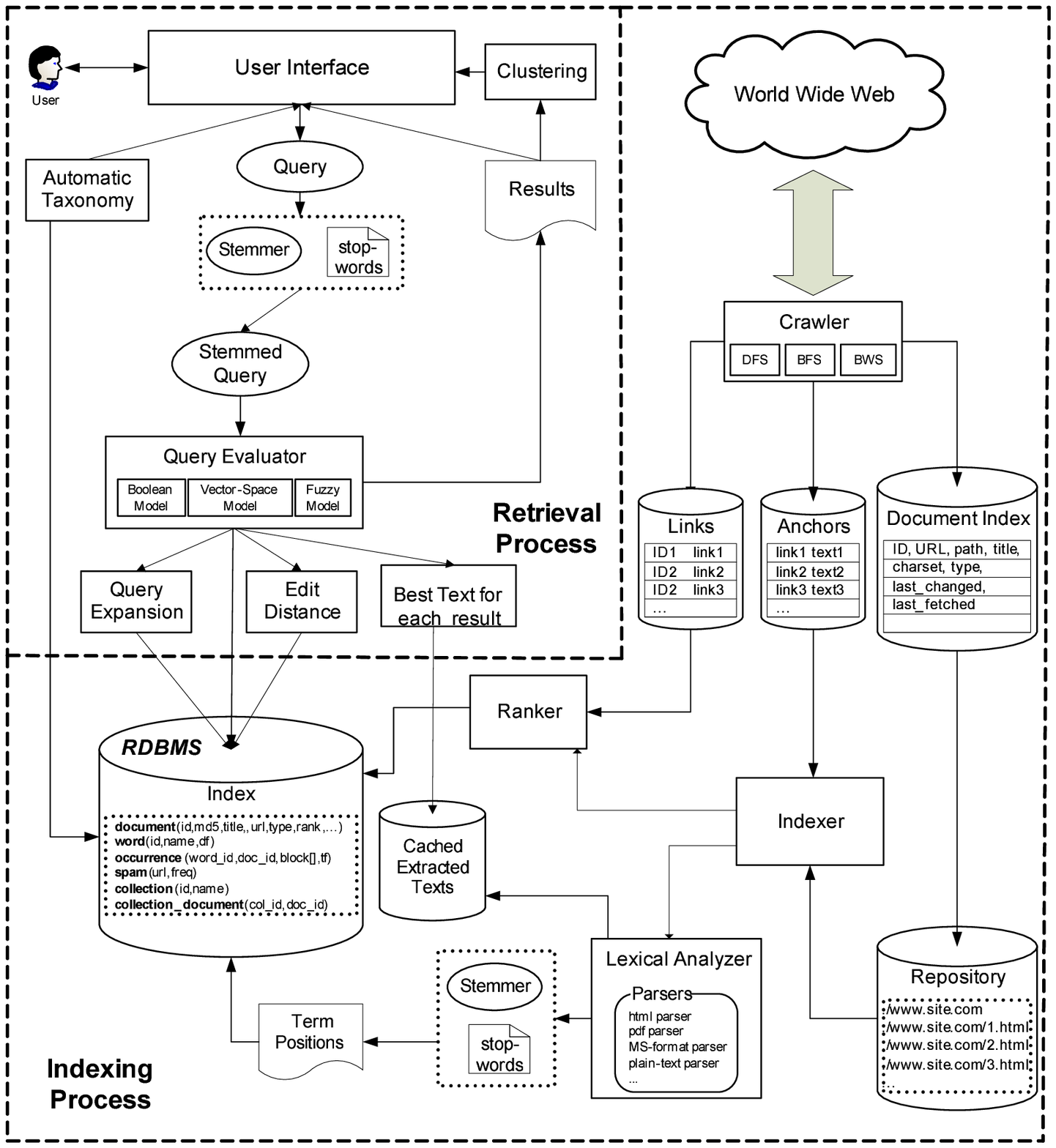}{115mm}{fig:groogleArchitecture}
  {The Component Model of \groogle}{}

At the indexing phase, the {\em crawler} is responsible for
recursively fetching Web pages given a starting
list of URLs to visit. 
The fetched pages are stored in a local {\em repository} for further processing.
Each downloaded page is assigned
a  unique ID number
and that number is used for referring to that page (e.g. in  {\em Links}).
The crawler
also builds a {\em Document Index}
that maps the ID of each page with its url
and other useful  properties
(like path, title, last changed/fetched, etc).
In addition, it stores the hyperlinks (including the anchor texts)
in the file {\em Links}.
This file
is useful
for  various
link analysis-based ranking techniques
(like PageRank).
The contents of the pages
in the repository
are indexed by the {\em Indexer}.
At first,
the {\em Lexical Analyzer}
identifies tokens,
removes stopwords
and applies stemming algorithms.
Then documents are seen as  vectors of terms
and the {\em Index} is created using a relational DBMS.
For each document,
its word occurrences, their weights
and the PageRank of the document is stored.
The exact positions of each word occurrence in a document
can be optionally stored.
In addition, the full text of documents  is extracted and stored locally
(in order to enable the derivation of document surrogates while presenting
the results of retrieval)
The exact process is described later in Section \ref{sec:Indexer}.

The retrieval process starts after the submission of a user query.
Several retrieval models are supported (Section \ref{sec:QueryEvaluator}).
If a query term does not exist in the lexicon,
the system suggests replacing that term
with terms that exist and whose
{\em Edit Distance} (from the submitted term) is less than a threshold.
Moreover, for each query the system suggests
a list of terms which could be used to {\em expand} (refine) the submitted query
(these terms are derived by analyzing the  top-ranked documents).
Regarding the presentation of the search results,
each document is presented by a surrogate
that includes  a short indicative excerpt
(containing as much of the query terms as possible).
In addition, search results are optionally {\em clustered} into different groups, so that each
group share some common trait.
Various clustering algorithms are supported (Section \ref{sec:Clusterer}).

\groogle\ offers a number
of auxiliary modules
including
a module for automatically constructing taxonomies
(Section \ref{sec:Taxonomy}) and a module
for aiding the detection of spam pages
in particular it identifies the subset of the fetched pages
that are worth inspecting by a human
(the pages that are manually marked as spam,
are then taken into account by the link analysis-based ranking techniques).

%

\section{Components Description}
\label{sec:Components}

This section describes each component in more detail.
Table \ref{tbl:Notations} shows some notations that will be used in
the sequel.

\puttableSV{tbl:Notations}{Notations}{}{
{\footnotesize
\btbl{|c|l|}\hline
Symbol & Meaning \\\hline\hline
$N$     & Number of documents indexed at the repository\\\hline
$E$     & Number of web links at the repository \\\hline
$K$     & Edit distance tolerance \\\hline
$L$ & 	Number of top pages used in query expansion and clustering,\\
    &   levels of automatic taxonomy\\\hline
$M$ & Levels with highest frequencies in automatic taxonomy\\\hline
$S$ & Number of query expansion suggested terms \\\hline
$V$     & Vocabulary\\\hline
$q$     & query \\\hline
$|q|$   & number of words in the query \\\hline
$|ans(q)|$ & number of documents with not empty degree of relevance with $q$ \\\hline
\etbl
}
}
Regarding experiments,
we crawled  all documents
of
our university department (http://www.csd.uoc.gr)
and FORTH-ICS (http://www.ics.forth.gr) web sites.
The whole collection is approximately 2.8 GB. About
32\% of the files use a Greek character encoding.

\subsection{Crawler}
\label{sec:Crawler}

The crawler roams the web, identifies all the hyperlinks
in each page and adds them to a list of URLs to visit.
URLs are then recursively visited according to a set of
policies. Currently, three traversal policies are supported:
Breadth-first (BFS), Depth-first (DFS) and Depth-within-site (DWS).
Crawler can be configured to download only files of a
specific type (e.g. html, pdf, rdf) as well as to ignore
others based on extension (e.g. *.tmp). The
identification of files is based on extension and on
content for dynamic web pages. Furthermore it is
compatible with the Robots Exclusion Protocol\footnote{
    http://en.wikipedia.org/wiki/Robots.txt
} to ignore
specified files or directories. The downloaded documents
are stored in the local file system. Each web site
occupies a different directory that has the name of its
domain (e.g. http://www.cnn.com will be stored under the
directory ./www.cnn.com) and its documents are stored
locally under the same absolute paths. A Document Index,
created also by the crawler, keeps information about
each document. These include the md5 of the document,
its canonicalized URL, the absolute path in the disk,
its title, its type (e.g. html, pdf, etc.) and the dates
that was last modified and last fetched. Furthermore, the
out-links of each downloaded web page are stored. The md5
of the document is computed based on its URL to eliminate
the possibility that two documents will have the same hash.
We have not observed any collision so far.
The hash is used as a unique identifier (ID) for the
document.
The Document Index is used by the Indexer to analyze
all downloaded documents and build the index.
To overcome the time spent due to I/O
latencies, the crawler uses multiple threads. Table
\ref{tbl:BFS-DFS} shows the time needed to
download 100.000 documents from popular servers all over the
world regarding the number of threads for BFS and DFS
algorithms. Each thread uses one distinct site as a seed, so
different threads do not overlap. The efficiency is almost
the same for both algorithms. We also observe a linear
speedup for the number of threads that is, however, highly
dependable on the selected web servers and the link capacity.

\puttableSV{tbl:BFS-DFS}{Times to crawl and download 100.000 pages}{}{
{\footnotesize
\btbl{|c||c|c|}\hline
Number of  & \multicolumn{2}{|c|}{Traversal Algorithm} \\\hline
Threads   & BFS & DFS \\\hline\hline
5      & 13552 sec & 13751 sec\\\hline
10     & 6725 sec & 6687 sec\\\hline
20     & 3562 sec & 3763 sec\\\hline
\etbl
}
}

%
%
%
%

\subsection{Lexical Analyzer}
\label{sec:Lex}

The Lexical Analyzer plays a major part in the
pre-processing of the documents. It is responsible
for converting a string of characters into a stream
of tokens. Most IR systems use single words as terms.
The Lexical Analyzer is called by the indexer for each
document, with its file type and encoding as parameters.
After processing the document it returns a hash map that
contains all document's words, along with their frequency and position.

The process of document analysis can be divided in the following steps:
\begin{enumerate}
\item Recognition of the document's structure
\item Extraction of document's text. The text is analyzed in tokens and
the terms to be inserted in the lexicon are selected.
For each token $t_i$ identified in the document $d_j$,
the analyzer returns the frequency (i.e $freq_{ij}$ ), the maximum
frequency (i.e $max_k \{ freq_{kj} \; | \;  t_k \in d_j \})$)
of terms in the document and the positions of $t_i$ in
$d_j$. Then, the frequencies are normalized by the
frequency of the term that appears mostly in the document.
\item Insertion of stemmed terms into the hash map
\end{enumerate}



The selection of the terms to be used as index terms
determines the quality of the lexicon. Words that are
considered non-informative, like function words
(the, in, of, a) also called stop-words are often ignored.
There is a list of stop-words which consists of 400 words
(320 english and 80 greek words).
The main benefit of stop-words removal
is that it reduces the size of the index,
however it may reduce recall
(e.g. consider a user is looking for documents containing the phrase
'to be or not to be').
For this reason, some web
search engines do not eliminate stop-words from documents.
In \groogle, stop-words elimination is optional.
Furthermore, there are options for the removal of numbers and
terms consisted of both characters and numbers.

The lexical analyzer accepts the following file types: html
(html, htm, php, jsp, asp), doc, ppt, pps, xls, rtf, txt.
For the text extraction from the documents various
software components were used,
specifically
{\tt pdfbox}\footnote{
    http://www.pdfbox.org/
}
for pdf documents,
{\tt Jakarta POI}\footnote{
    http://poi.apache.org/
}
for doc, ppt, pps, xls,
and {\tt RTFEditorKit}  for RDF documents.
The analyzing time of each document type is shown in Table \ref{tbl:AnalyzingTime}.

\puttableSV{tbl:AnalyzingTime}{Lexical Analysis Times}{}{
\btbl{|c|c|}\hline
File Type   &   Average Time of 100 experiments (sec)   \\\hline
html        &   0.0361                 \\\hline
pdf     &   0.3750                 \\\hline
doc     &   0.0808                 \\\hline
ppt     &   0.0732                 \\\hline
pps     &   0.2630                 \\\hline
xls     &   0.0983                 \\\hline
rtf     &   0.1917                 \\\hline
txt     &   0.0343                 \\\hline
\etbl
}


\subsection{Stemmer (including a Greek one)}
\label{sec:Stemmer}

The main idea behind stemming is that users searching for information on
\textit{retrieval} will also be interested in articles that have information
about \textit{retrieve}, \textit{retrieved}, \textit{retrieving},
\textit{retriever}, and so on.
Several stemmers for various languages have been developed over the years,
each with its own set of stemming rules. However, the usefulness of stemming
for improved search quality has always been questioned in the research
community, especially for English. The consensus is that, for English,
stemming yields small improvements in search effectiveness; however,
in cases where it causes poor retrieval, the user can be considerably
annoyed \cite{hull96stemming}.

Stemming is possibly more beneficial for languages with many word
inflexions (like German and Greek) \cite{Nakov}. Currently there were only few
stemmers for the Greek language \cite{Ntais}. The stemmer used in \groogle\ is based
on the automated technique of affix removal - Porter's Algorithm \cite{porter}.
By studying the greek language grammar rules,
a collection of common suffixes for nouns and adjectives was
gathered, including comparative, participle, singular and plural
forms in all genders and also verbs in different voices and tenses.
This collection was expanded using 270 "productive suffixes" for
words produced by other words with the same semantics. The total
number of suffixes used is 782. In the simple
past and past continuous tenses the inflexion may be applied with a
possible addition of an accentuated letter as a prefix, or internal
change in case of compound words. Therefore a collection of adverbs
and prothesses as prefixes was considered necessary to eliminate
these inflections and derive the base word of the compound. Another
point where the stemming could be problematic when using common methods
of affix removal, are the irregularities of verbs. The restricted set
of these words was used to reduce the semantically equal but lexically
different roots into a unique root. Also a collection of unmodified
words was gathered, since the stemming process would be useless.
Finally, the inflexions of the last character of the root, provide
according to ancient greek grammar rules, three different sets of
characters and digraphs and a single character which will replace
each of these sets.

The greek stemmer is initialized, using the files of suffixes,
prefixes, irregular verbs and unmodified words to construct an equal
number of Trie structures. The prefixes file contains 83 entries and
as far as the irregular verbs is concerned, variations of 29 different
verbs were used. Additional information is kept in
the suffixes trie, reffering to the type, verbal or not, of the word
that is matched. Also in the prefixes trie the initial form is kept
as a replacement. The irregular verbs trie contains the stemmed
variations and a predefined stem.

\puttableSV{tbl:GStemExamples}{Examples of Greek Stemmer}{}{
{
\scriptsize
\btbl{|c|c|c|c|c|} \hline
Word & Word Split & prefixes-First Stem & Increment-Alternate & Final Stem \\\hline $\pi
\rho \alpha \tau \tau \omega $ & $\pi \rho \alpha \tau \tau \omega $
& $\pi \rho \alpha \tau \tau $ & $\pi \rho \alpha \xi $ & $\pi \rho
\alpha \xi $
\\\hline $\pi \rho \alpha \kappa \tau \iota \kappa o \varsigma $ &
$\pi \rho \alpha \kappa \tau \iota \kappa\varsigma $ & $\pi \rho
\alpha \kappa \tau $ & $\pi \rho \alpha \xi $ & $\pi \rho \alpha \xi
$ \\\hline $\pi \rho \alpha \xi \eta $ & $\pi \rho \alpha \xi \eta $
& $\pi \rho \alpha \xi $ & $\pi \rho \alpha \xi$ & $\pi \rho \alpha
\xi$ \\\hline $\pi\rho\alpha\gamma\mu\alpha$ &
$\pi\rho\alpha\gamma\mu\alpha$ & $\pi\rho\alpha\gamma\mu$&
$\pi\rho\alpha\xi$ & $\pi\rho\alpha\xi$
\\\hline $\alpha\nu\alpha\delta\iota\alpha\tau\alpha\xi\eta$ & $\alpha\nu\alpha - \delta\iota\alpha - \tau\alpha\xi\eta$ & $\alpha\nu\alpha - \delta\iota\alpha - \tau\alpha\xi$ & $\alpha\nu\alpha - \delta\iota\alpha - \tau\alpha\xi$
& $\alpha\nu\alpha\delta\iota\alpha\tau\alpha\xi$
\\\hline $\alpha\nu\alpha\delta\iota\alpha\tau\alpha\sigma\sigma\omega$& $\alpha\nu\alpha - \delta\iota\alpha - \tau\alpha\sigma\sigma\omega$ & $\alpha\nu\alpha - \delta\iota\alpha - \tau\alpha\sigma\sigma$
& $\alpha\nu\alpha - \delta\iota\alpha - \tau\alpha\xi$ &
$\alpha\nu\alpha\delta\iota\alpha\tau\alpha\xi$
\\\hline $\alpha\nu\alpha\delta\iota\epsilon\tau\alpha\xi\alpha$& $\alpha\nu\alpha - \delta\iota\epsilon - \tau\alpha\xi\alpha$ & $\alpha\nu\alpha - \delta\iota\alpha - \tau\alpha\xi$
& $\alpha\nu\alpha - \delta\iota\alpha - \tau\alpha\xi$ &
$\alpha\nu\alpha\delta\iota\alpha\tau\alpha\xi$
\\\hline $\pi\alpha\omega$ & $\pi\alpha\omega$ & $\pi$ & $\pi\eta\gamma$ & $\pi\eta\gamma$ \\\hline $\pi\eta\gamma\alpha\iota\nu\omega $ & $\pi\eta\gamma\alpha\iota\nu\omega $ & $\pi\eta\gamma$& $\pi\eta\gamma$&$\pi\eta\gamma$ \\\hline
 \etbl
}
}

A word to be stemmed is first checked whether it belongs to an unmodified words set.
Currently, this set contains 291 words. The actual stemming
procedure starts with the prefixes separation which produces
alternative prefixes. Suffix removal is applied on the
derived word after the replacement of all accented-characters. Since
the current stem comes from a verbal suffix type, it is examined if it
belongs to the irregular verbs in order to retrieve the replacement
stem. In other cases an optimization is conducted. The optimization
may increase the length of the root by one or two letters and may be
followed by a last character replacement. Finally, existing prefix
is reconcatenated to produce the final stem of the word. Examples of the process
can be seen in Table \ref{tbl:GStemExamples}.

\subsection{Lexicon}
\label{sec:Lexicon}

We conducted some measurements in order
to see the distribution of words in our collection
and the effects of stemming and stop-words removal.

Figure \ref{fig:words_freq}a shows the frequencies
of terms in the Index, when stemming is not enabled and stopwords have not been removed.
One can observe that a
few terms, including many stop-words (like the terms 'and',
'the', 'for', 'with', etc.) get upward of 4000 occurrences, whereas most
terms got only a few occurrences (about 54\% of the terms
have got only a single occurrence). Specifically the index has 301052 terms with 4824739 occurences.
Figure \ref{fig:log_words_freq}a shows the same plot,
but on a log-log scale.

To investigate to what extent
the plotted function approximates a power law,
we rely on a commonly used method
(based on the least square errors method),
called Linear Regression \cite{linReg}, to
fit a line in a set of 2-dimensional points and, thus,
to investigate
whether the \emph{log-log} plot of the function approximates a line.
The accuracy of the approximation
is indicated by the correlation coefficient,
the absolute value of which (hereafter called $ACC$)
always lies in $[0,1]$. An $ACC$ value $1.0$ indicates perfect
linear correlation, i.e., the points are exactly on a line.
In this respect,
we observed that
the function plotted in Figure \ref{fig:words_freq}
approximates ($ACC = 0.996$, i.e. $99.6\%$ confidence)
a power law with exponent $1.29$.

\putTwoFiguresEngl{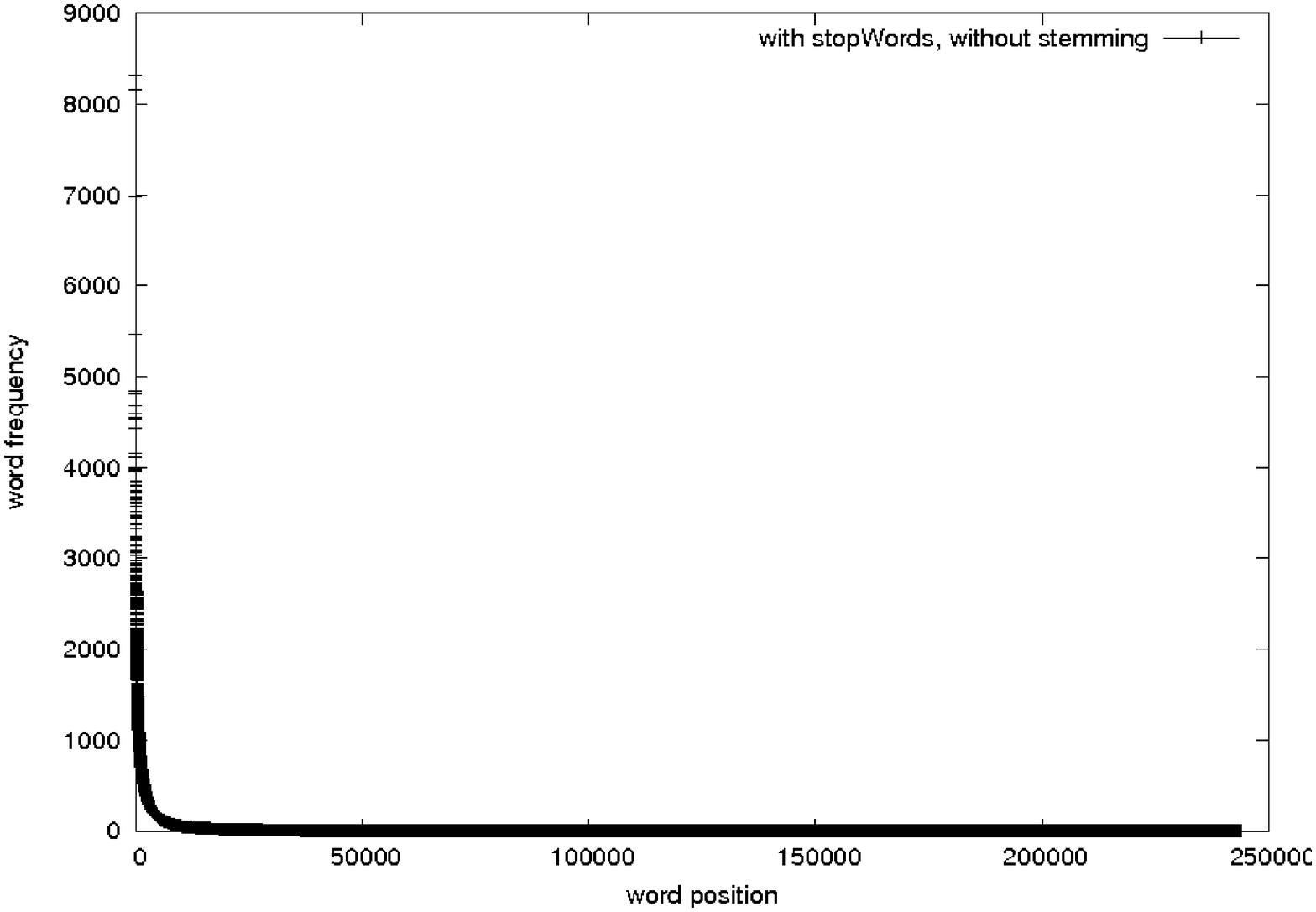}{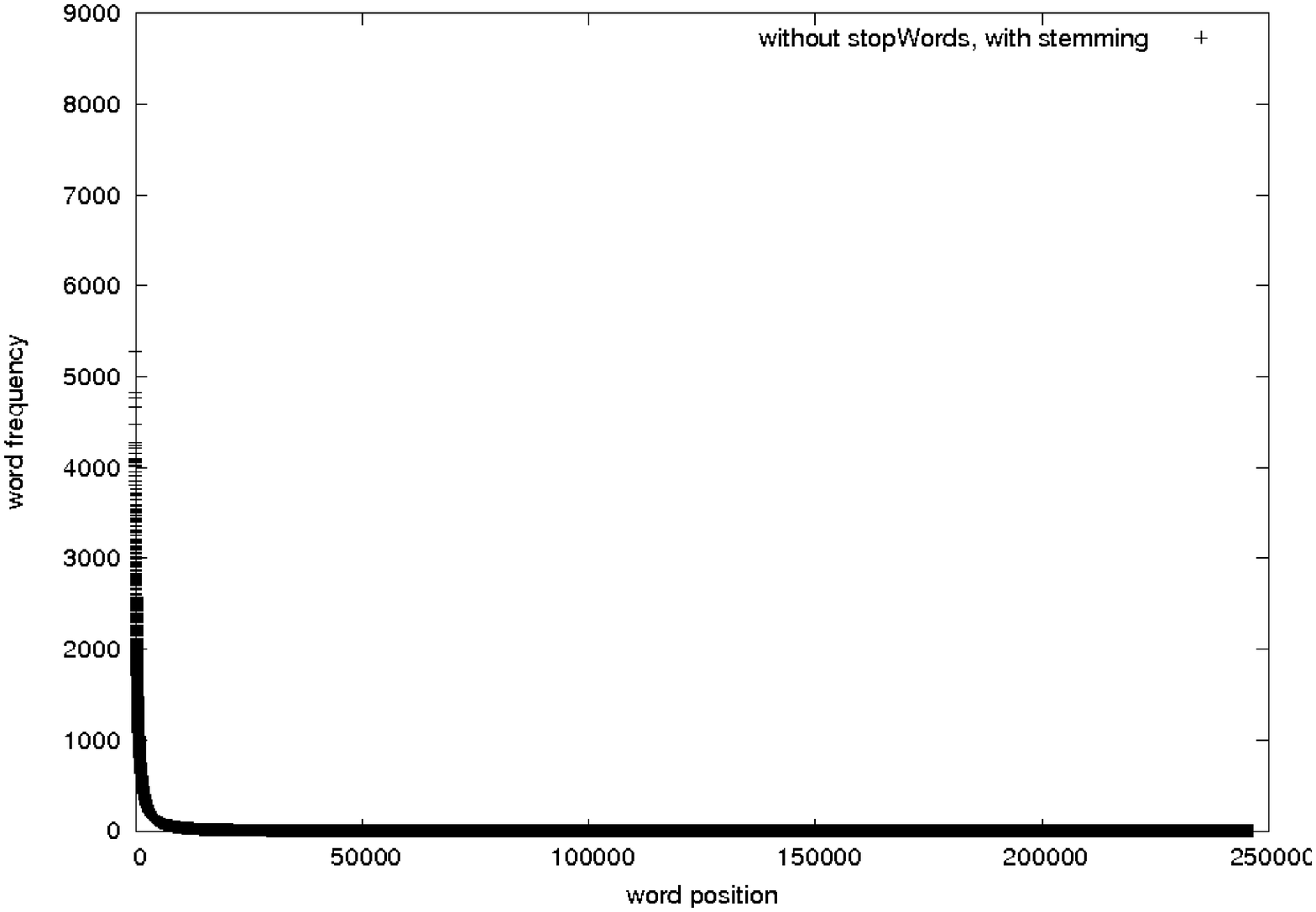}{35mm}{fig:words_freq}
{Linear scale plot of the words and their frequencies. In the right plot the words are stemmed and stop-words have been removed while in the left not.}{}

Figures \ref{fig:words_freq}b and
\ref{fig:log_words_freq}b show the term frequencies
when stemming is enabled and stop-words have been removed.
Now our index consists of 220518 terms (26.7\% reduction caused by stemming) and 3435040 occurences (28.8\% reduction caused by stopwords).
That function also
approximates ($ACC = 0.996$)
a power law but with
slightly decreased exponent, i.e. $1.18$.

Although the \emph{log-log} distributions of both functions
follow a power law, we observe a top concavity deviation,
frequently met on many datasets\cite{Faloutsos06}.






\putTwoFiguresEngl{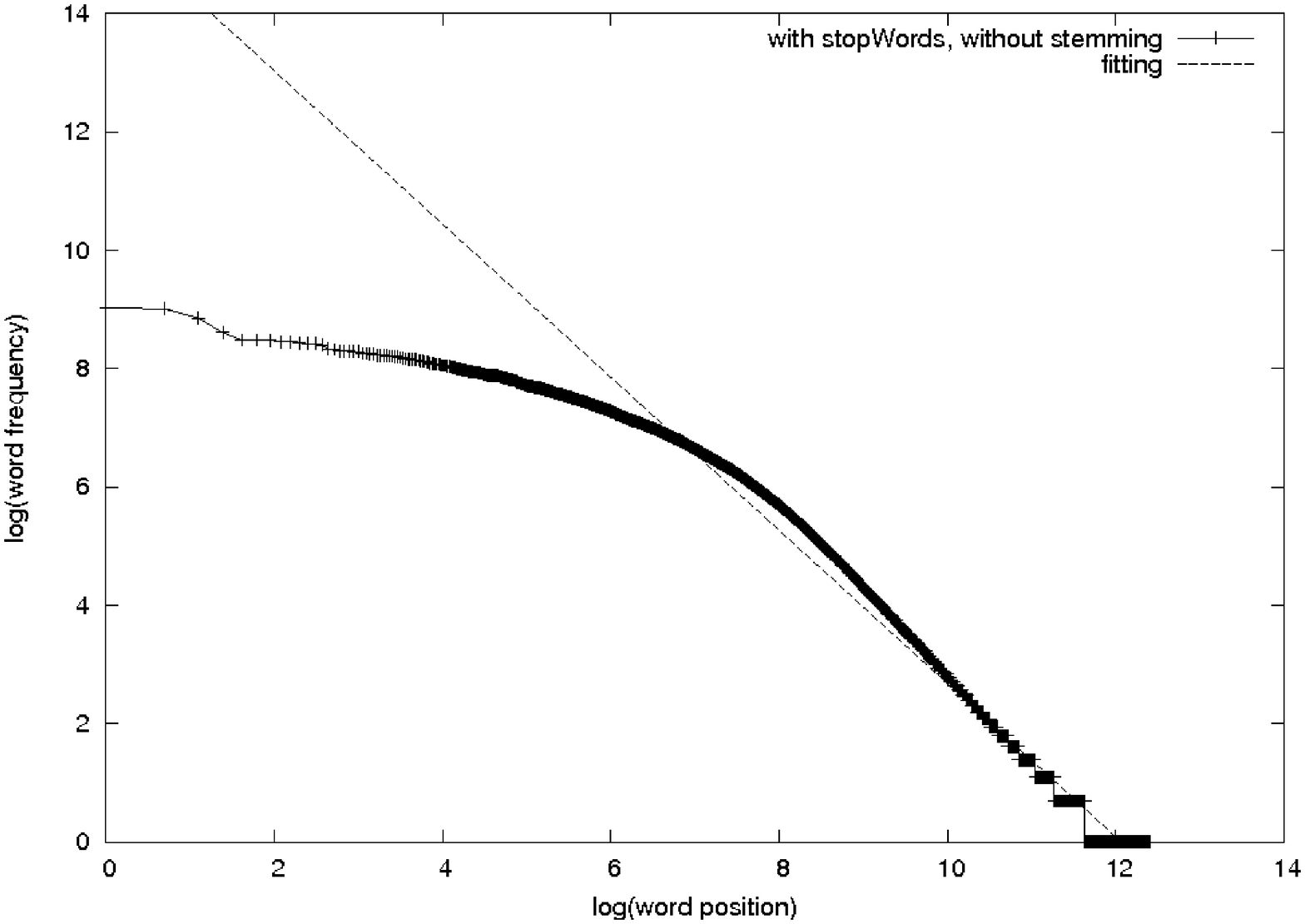}{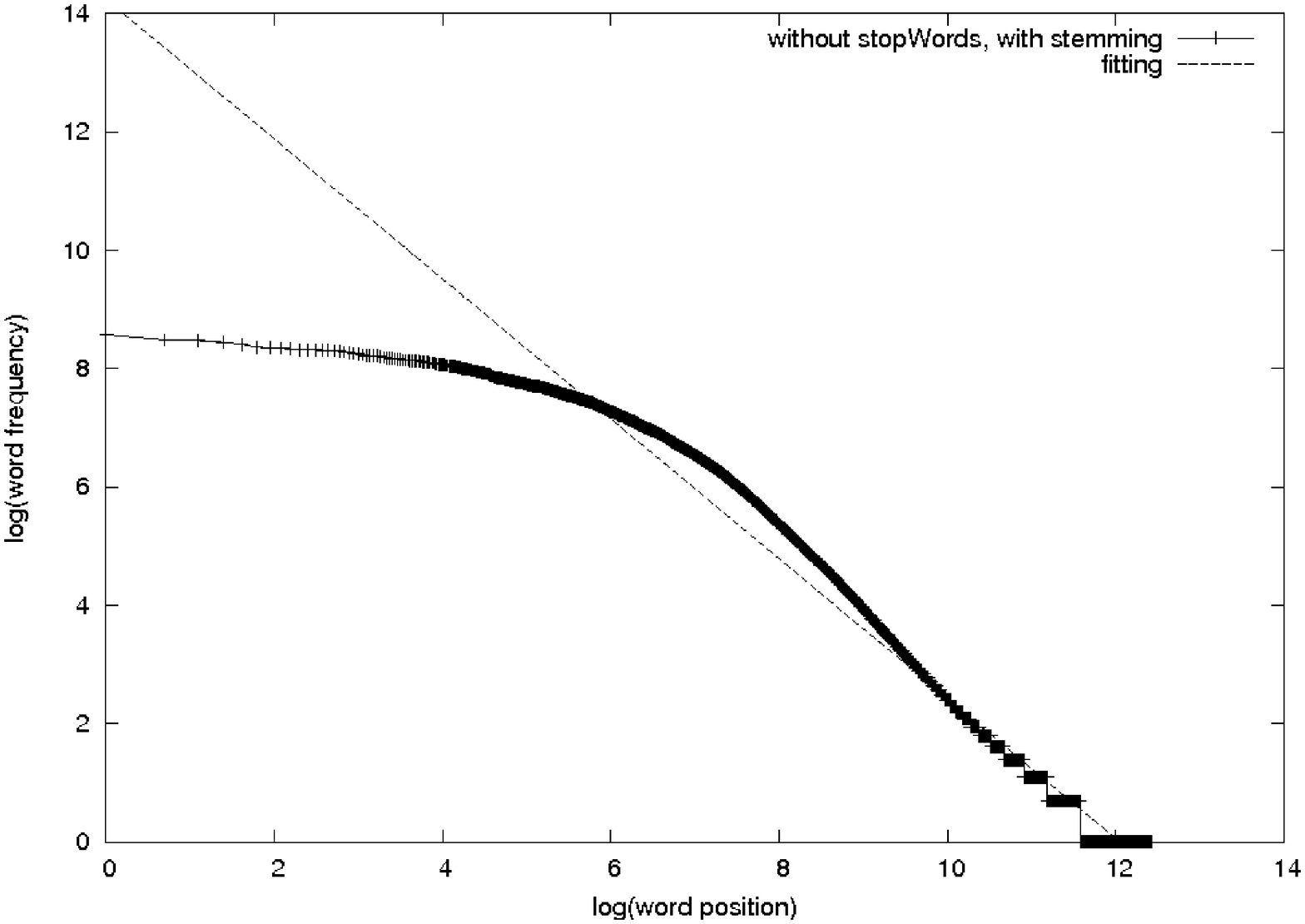}{35mm}{fig:log_words_freq}
{Log-log scale plot of the words and their frequencies. In the right plot the words are stemmed and stop-words have been removed while in the left not.}{}

\subsection{DBMS-based Indexer}
\label{sec:Indexer}

The Indexer iterates through all the records of the
Document Index and uses the Lexical Analyzer component
to create a hash table that contains the words and
their exact positions for each document in the
Repository.
The index is built on top of a DBMS
(in particular over PostgreSQL 8.3).
The database schema can be seen in Table \ref{tbl:schema}.

The use of a relational DBMS is motivated by the
following facts:

\bi
    \item[i)]
    it allows for incremental indexing of document collections.
    To the best of our knowledge,
    no other open-source IR system
    allows incremental indexing,
    i.e., one should always create the index from scratch.
    Furthermore, the "deletion" of documents from the index
    (e.g. when a Web page disappears or when it is classified as spam)
    is efficient,
    while this operation is very expensive in  inverted files\footnote{
      The cost of this operation in an inverted file is $\cal{O}$$(n)$ where $n$ is the size of
      the text
      of the entire collection.
    }.
    Savings in time and computational resources are straightforward.
    Moreover, this functionality is especially useful,
    for the IR research community,
    which frequently faces the need
    to conduct experiments on top of an IR system.
    \item[ii)]
    makes our index to exhibit the well-known ACID properties.
    In this manner,
    the index can be appended by an new document collection
    while at the same time being able to respond to user queries.
    \item[iii)]
    allows for efficient information retrieval by the
    advanced query planing and optimization features.
    Such capabilities are very useful
    when ranking is based on  complex ranking formulas
    (current IR systems are commonly optimized only for one kind of queries).
    \comment{
        In this respect we take advantage
        of  thirty and more years of optimization
        by the DBMS community.
    }
\ei

In addition,
the use of PostgreSQL among other relational DBMSs
is justified by the fact that it
supports a family of (easily extensible)
secondary memory indices, called SP-GiST~\cite{SPGiST1,SPGiST2},
that allow for advanced indexing functionality.
For instance,
the well-known structure of Tries,
which has been extensively used by IR systems,
has been implemented~\cite{ToTrieOrNotToTrie} as a member
of the more general category of SP-GiST indices. According to
\cite{Eltabakh06}, it offers more than 150\% performance increase for
exact search matches over to postgreSQL B+-trees, and scales better especially
with the increase in the data size.
Another,
worth studying index that has been built on top of PostgreSQL
is the tree-Trie index,
appropriate for indexing relationships
with set-value attributes~\cite{treeTrie}.
Although at the current state
of \groogle\ implementation,
we have not yet experimented
with such advanced PostgreSQL indices,
their study seems a promising direction
for our future research and development efforts.

\puttableSV{tbl:schema}{Database schema of the Index. In parentheses is shown the size (in bytes) of each field.}{}{
{\scriptsize
\btbl{|l|l|l|}
\hline
$Table  $   & $Field$           & $Type$ (Bytes)  \\
\hline
document    & id                & $\mathsf{int4}$  (4)         \\
            & md5               & $\mathsf{char}$ (16)      \\
            & title             & $\mathsf{varchar}$ (title.length)   \\
            & path              & $\mathsf{varchar}$ (path.length)     \\
            & link              & $\mathsf{varchar}$ (link.length)  \\
            & type              & $\mathsf{varchar}$ (type.length)     \\
            & encoding          & $\mathsf{varchar}$ (encoding.length) \\
            & norm              & $\mathsf{float}$   (4)  \\
            & rank              & $\mathsf{float}$   (4)  \\
\hline
word        & id                & $\mathsf{int4}$ (4)\\
            & name              & $\mathsf{varchar}$ (name.length)\\
            & df                & $\mathsf{int4}$ (4)\\
\hline
occurence   & word\_id          & $\mathsf{int4}$ (4)\\
            & doc\_id           & $\mathsf{int4}$ (4)\\
            & block[]           & $\mathsf{Array \langle int4 \rangle}$ (4$\times$ block.length)\\
            & tf                & $\mathsf{float}$ (4)\\
\hline
spam        & url               & $\mathsf{varchar}$ (url.length)\\
            & freq              & $\mathsf{int4}$ (4)\\
\hline
collection  & id                & $\mathsf{int4}$ (4)\\
            & name              & $\mathsf{varchar}$ (name.length) \\
\hline
collection\_document
            & col\_id(4)        & $\mathsf{int4}$ (4)\\
            & doc\_id(4)        & $\mathsf{int4}$ (4)\\
\hline
\etbl
}
}

The {\tt document} table keeps all necessary
information for each document. The Lexicon is stored
in the {\tt word} table. The current Lexicon contains about 250
thousand words. A separate table is used to store
the word offsets. We take advantage of the array data type
that is supported by PostgreSQL, to avoid the insertion of
a new tuple for every occurrence - all occurrences of a
word in a document will be in the same tuple.
In this respect,
we exploit the idea of the inverted index,
by modeling the term occurrences in documents
as a set-value attribute
(e.g., array of integers),
while at the same time
relying on a relational DBMS.

Indexer can optionally use block addressing in order to
reduce space requirements for storing term positions. The
occurrences of a term that are near can be grouped
together proportional to the block size. Block size can
either be fixed, so each document will have a variable
number of blocks, or can vary, so that the number
of blocks will be the same for each document.

Of course we can further reduce the space requirements
of our index, by storing only the \emph{tf} (normalized) of a term
in a document and not its occurrences.
Hence, we drop the attribute {\tt block} of table {\tt occurence}.
In that case,
when \groogle\ should report the query results to the user,
we have to parse the text file of a document to find the
best text with respect to the query terms, because this information
does not exist in the index.
This is the approach adopted by
the current \groogle\ working installation.

\subsubsection{Bulk Index Creation/Updates}\label{sec:IndexerCreUpd}

At a first glance,
it seems that the benefits of the use of a relational DBMS
is at the expense of efficiency
of the data storage and retrieval.
For instance the guarantee
of the ACID properties implies
an additional time cost.
The concurrency control,
the update of DBMSs indices
(e.g., B-trees etc.)
and their possible reorganization on disc
due to the insertion of
new tuples in the corresponding relations,
are only two of the factors
which may harm the efficiency of our index.

In order to reduce the effect of these problems, we:
\bi
    \item[i)]
    use the copy function of PostgreSQL.
    In this manner, we
    skip the concurrency control,
    as well as several integrity constraints checks,
    while at the same time we minimize the I/O's needed
    to insert a specific amount of new tuples.
    \item[ii)]
    drop DBMSs indices on relations,
    afterwards index, and at the end
    re-create the same DBMSs indices.
    In this manner,
    we pay time only to compute the indices organization
    that corresponds to the final state of the relations contents,
    instead of paying time to compute temporal indices organizations
    that will need to be changed after the next tuple(s) insertion.
    \item[iii)]
    provide hints to the PostgreSQL query optimizer
    to force it to choose the optimal access paths
    (i.e., by taking advantage from the built relation indices)
    as well as the optimal query execution plan.
\ei

After all documents in the collection have been indexed,
for each document $d$
we compute the  norm ($\|\vec{d}\|$) of
its vector ($\vec{d}$)
as defined by the tf-idf weighting scheme,
and store it
in the {\tt norm} field.
This will speed-up the evaluation of a query at the
searching phase, as we will see in \ref{sec:QueryEvaluator}.

Furthermore, we also consider anchor texts at weighting since
they can provide better quality results \cite{Brin98}.
Anchor texts and the pointed URLs are stored offline
by the crawler. When indexing finishes, the terms of the anchors are
stored in the DBMS as if they were contained in the pointed document.
Thus, anchor terms are
affecting document weighting indirectly by increasing the
frequencies of the terms in the Lexicon.
If an anchor term does
not exist in the {\tt occurrence} table,
then the term is inserted in the table with a
{\tt tf} value equal to 0.5,
otherwise its {\tt tf} value is updated
using the formula $newTF = (oldTF + 0.5) / 1.5$.

To improve performance and increase the speed during
searching, we took advantage of indexing options offered by the DBMS.
In particular,
5 indices have been built:
a unique btree index on {\tt word.name},
a unique btree index on {\tt word.id},
a unique btree index on {\tt document.id},
a btree index on {\tt occurrence.word\_id}
and a btree index on {\tt occurence.doc\_id}. We have also
clustered {\tt occurrence} table according to
{\tt doc\_id} field, {\tt word} table according to {\tt name} field
and {\tt document} table according to {\tt id} field.
The last two tables of the
relational schema allow having  collections (like
News, GreekWeb, EntireWeb, etc), where a document
may belong to more than one collection.

\comment{==
Whenever a document is indexed, many tuples should be
inserted in the above tables resulting an increased
insertion cost (due to concurrency control, etc. $<$-\textbf{(check
if this is correct)}). To further increase the insertions in
the database, we first write many tuples in a file and
then copy the file in the database, a technique known as
$batch\ insertions$. Using this only 3 DB-accesses are needed (Copy functions for
documents, words and occurences files) a
process which is about 14 times faster\textbf.
==}

Table \ref{tbl:Indexing} shows the sizes and the
times required to build the Index in
various configurations
(of stemming and stop-words removal),
assuming the collection described at beginning of this  section.
We observe that term positions
constitute a significant factor
for the size of the Index. Nearly 40\% of the whole
Index is devoted to term positions, even in the
case where block addressing is used.
Besides that, even in the worst case, where term
positions are stored and full indexing is performed,
the size of the final Index is only 13\% of the
original collection size.

The overall efficiency is satisfactory
as 2.8 GB are indexed in less than an hour
(more on comparison with other systems at Section \ref{sec:RW}).


\puttableSV{tbl:Indexing}{Indexing}{}{
{\scriptsize
\btbl{|l|l|l|l|}\hline
$\textbf{Options}$  &   $\textbf{Index\ Size}$  &   $\textbf{\%\ Of\ Collection}$   &   $\textbf{Indexing\ Time}$  \\\hline
\multicolumn{4}{|c|}{\textbf{Without Term Positions}}                   \\\hline
Full                & 214.99 MB & 7.4\% & 2892 sec          \\\hline
Stop-words removed      & 195.06 MB & 6.8\% & 2808 sec          \\\hline
Stemmed words           & 174.54 MB & 6\%   & 2836 sec          \\\hline
Stop-words removed and stemmed  & 156.39 MB & 5.5\% & 2752 sec          \\\hline
\multicolumn{4}{|c|}{\textbf{With Term Positions (Block addressing - block size=2K)}}   \\\hline
Full                & 353.13 MB  & 12.3\%   & 3030 sec          \\\hline
Stop-words removed      & 315.64 MB  & 11\% & 2961 sec          \\\hline
Stemmed words           & 290.96 MB  & 10.1\%   & 2890 sec          \\\hline
Stop-words removed and stemmed  & 256.46 MB  & 8.9\%    & 2844 sec          \\\hline
\multicolumn{4}{|c|}{\textbf{With Term Positions}}                  \\\hline
Full                & 376.65 MB & 13.1\%    & 2943 sec          \\\hline
Stop-words removed      & 331.38 MB & 11.6\%    & 2883 sec          \\\hline
Stemmed words           & 317.08 MB & 11.1\%    & 2904 sec          \\\hline
Stop-words removed and stemmed  & 274.54 MB & 9.6\% & 2760 sec          \\\hline
\etbl
}
}

\subsection{Link Analysis-based Ranker}
\label{sec:Ranker}

The Ranker provides a number of link analysis techniques.
At first it constructs a
directed graph where each node represents a
fetched document and the edges of each node represent the
corresponding hyperlinks of that document.
The graph is constructed using the IDs and
the out-links of the fetched documents
that are stored in the Document Index (derived by the Cralwer).
It implements the PageRank \cite{Brin98} ranking algorithm
and the resulting ranks
are
stored in the {\tt rank} field of the {\tt document}
table in the Index.
Biased PageRank is also supported.
In particular,
the Ranker can receive as input a file that
may contain (a) spam pages and (b) preferred pages.
With such an input it behaves like a biased PageRank algorithm,
specifically like the TrustRank algorithm described in
\cite{TrustRank04}.
\comment{
    The PageRank of each web page is computed using the type
    \[
    PR(A) = (1-damp) + damp*(PR(T1)/C(T1) + ... + PR(Tn)/C(Tn))
    \]
    where A is the web page to be ranked, T1,T2,...,Tn are the
    web pages that point to A (inlinks), damp is a parameter
    with value between 0 and 1 and C(A) is the number of external
    links of A.
}
To avoid leaking edges on the web graph, we consider only the
out-links that point to documents that have been fetched.
We do not want to rank pages that are not included in the
repository. We treat those pages as if they were spam pages
and as a consequence they do not participate in rank
computation.

To combat spam pages,
it is important to be able to identify
pages that are well connected with
other pages.
If such a well-connected page is a spam page then it will
transport its rank
to a lot of other (probably spam) pages.
To identify such pages the approach described in
\cite{TrustRank04} is followed,
i.e. we compute the Inverse PageRank
(whose computation is similar to PageRank by reversing the
direction of edges).
The top ranked pages are then proposed for human inspection.

\puttableSV{tbl:RankTimes}{Execution times for ranking algorithms}{}{
\btbl{|c|c|c||c|c|c|}\hline
$N$ & $E$ & $i$ & $PageRank$ & $Biased\ PageRank$ & $Inverse\ PageRank$\\\hline\hline
1000    &   23175   &   10  &   0.0503 &   0.0567 &   0.0660\\\hline
2000    &   36052   &   10  &   0.0798 &   0.0856 &   0.1025 \\\hline
3000    &   54115   &   10  &   0.1187 &   0.1296 &   0.1612 \\\hline
10000   &   133510  &   11  &   0.3236  &   0.3562 &   0.4294 \\\hline
14723   &   151616  &   11  &   0.3937 &   0.4184 &   0.5032 \\\hline
\etbl
}

Table \ref{tbl:RankTimes} reports the execution times for
running PageRank, Inverse PageRank and Biased PageRank
and stopping after $\log N$ iterations, where $N$ denotes
the number of pages and $E$ the number of links.
We observe that the execution times are quite fast\footnote{
This means that we could probably offer
biased PageRank, for personalization purposes, at realtime.
}.

Also note that the stored ranks can be exploited by the crawler,
so that pages with higher rank will be downloaded earlier.

\subsection{Query Evaluator}
\label{sec:QueryEvaluator}

The Query Evaluator allows users to  restrict the
search space according to file type of the
sought documents (e.g. .html, .pdf, .ps, .ppt, .doc).
In addition, it supports several retrieval models,
specifically the
Vector Space (VSM),
the Boolean, the Extended Boolean,
and the Fuzzy Model (under  $tf*idf$ weighting as described in \cite{Baeza99}).
The last three allow user queries to contain AND, OR, and NOT operators.
Apart from this, user queries are treated
like the indexed documents,
so stemming and stopwords removal can optionally be applied on them.
In case any of the query terms do not exist
in the index, \groogle\ proposes alternatives to the user,
by using the {\em Edit distance} algorithm,
searching the index for those words whose distance is
less or equal than a predefined constant $K$ (default = 2) and whose document frequency
is from the smallest ones, namely words with big discreet ability.
The user selects the desired term and
the corrected query is evaluated again.
To find the alternatives, the classical
dynamic programming algorithm
is used (whose time complexity is independent of $K$).
Table \ref{tbl:EditDistanceTimes1}
reports  the average time needed to find possible
matches to miss-spelled or incorrect words (for $K=1$
or more values) using our Edit Distance
implementation.
We used a set of 100
miss-spelled words that was created by doing
random transformations (insertion, deletion or
substitution) in words from the index.
We observe that the execution time is proportional
to the vocabulary size. Threshold $K$ does not
affect execution times, as expected.

\puttableSV{tbl:EditDistanceTimes1}{Execution times for Edit Distance}{}{
\btbl{|c|c|c|c|}\hline
$K$ &   \multicolumn{3}{|c|}{Average Time of 100 experiments (sec)} \\\hline
    &   $|V|=50000$ &   $|V|=100000$    &   $|V|=200000$    \\\hline
1   &   0.1450     &   0.2922      &   0.5895     \\\hline
5   &   0.1458     &   0.2938     &   0.5992     \\\hline
10  &   0.1438     &   0.2908      &   0.5861     \\\hline

\etbl
}

As the Boolean Model is an exact match model,
query results are ranked according to PageRank.
However, the  retrieval model that is used by default is a {\em hybrid model}
combining VSM and PageRank.
The formula that is currently used
for computing the similarity between a document an a query is:
\[
sim(d,q) = 0.7 * CosSim(d,q) + 0.3 * PageRank(d)
\]

A new hybrid retrieval model is under construction.
Let $|q|$ denote the number of words in $q$.
The answer of $q$, denoted by $answer(q)$
is a linear order of blocks,
i.e.
$answer(q) = \langle B_{|q|}, \ldots, B_1 \rangle$,
where $B_i$ comprises all those pages what have
$i$ words of the query ( $1 \leq i \leq |q|$).
The elements of each block
$B_i$ are ordered according to $sim(d,q)$ as defined earlier.

\subsection{Local Automatic Analysis for Query Expansion}
\label{sec:Expansion}

\comment{
    Pseudo-relevance feedback, also known as local feedback or blind
    feedback, is a tecnnique commonly used to improve retrieval
    performance \cite{QueryExp96},\cite{buckley-new}.
}
The initial query (as provided by the user) may be an inadequate
or incomplete representation of the user's information need.
The aim of this component  is to aid users in reformulating
their original query by suggesting additional terms.
In our case, the system suggests
terms that occur frequently in the top-ranked documents
of the answer of the original query.
These terms are selected using a  very simple and fast method.
For each term $t_i$ that appears in the
top $L$ (by default $L$=5)
documents returned by the Query Evaluator,
we sum its term frequencies (i.e. all {$tf_{ij}$} where $j$ in top-$L$ documents)
and we recommend to the user the $S$ terms (by default $S$=5) with the highest accumulative frequency.
Some examples
are shown in Table \ref{tbl:QueryExpansion}.
The computation is very fast and has almost linear time complexity
with respect to $L$ (depending of course to the number of different terms appearing in the top L documents).
Mean execution times with the same query, $S$=5 and different $L$ values are shown in Table
\ref{tbl:QueryExpansionTimes}.

\puttableSV{tbl:QueryExpansion}{Query Expansion Examples}{}{
{\footnotesize
\btbl{|l|l|l|l|l|l|l|}\hline
    &   $Initial\ Query$    &   \multicolumn{5}{|c|}{$Expanded\ Terms$} \\\hline\hline
1   & retrieval & imag & medic & index & storag & system    \\\hline
2   & web       & system & servic & page & process & cours  \\\hline
3   & user      & interfac & layer & system & develop & softwar \\\hline
\etbl
}
}

\puttableSV{tbl:QueryExpansionTimes}{Query Expansion Average Times}{}{
{\footnotesize
\btbl{|l|c|}\hline
 $L$   &  $Time\ (sec)$ \\\hline\hline
5      & 0.002          \\\hline
10     & 0.003          \\\hline
15     & 0.004          \\\hline
20     & 0.004          \\\hline
\etbl
}
}

\subsection{Result Clustering}
\label{sec:Clusterer}


Results clustering in Web searching is provided only by a limited number of  search engines
mainly because it is rather a computationally expensive task.
Vivisimo\footnote{www.vivisimo.com } probably offers the best results clustering today.
Our objective was to provide an efficient and effective method for results clustering.
As in Web searching the results of a query can be numerous,
hierarchical clustering
seems a more appropriate choice than flat clustering,
as it could give an overview of a large answer set.
Instead of applying a hierarchical agglomerative clustering algorithm,
\groogle\ offers several
more efficient algorithms all based on K-means.
Although K-means derives a flat set of clusters
we extended it
with an additional step
that assigns a name to each of  the clusters derived
and
methods for building  hierarchies over these names.
Let $C = \{C_1, \ldots C_K\}$ be the $K$ clusters
obtained by applying K-means algorithm over the set of top-$L$ documents of the current answer
($L$ is a parameter, whose default value is 100).
For each $i=1 \ldots K$ we use $c_i$ to denote the centroid vector of $C_i$.
The set of names of a set of clusters $C$
 comprises the $m$-most weighted terms of each $c_i$
where $m$ is the  smallest integer
such that all clusters have a distinct name (i.e. a distinct set of words).
Below we just sketch the methods that are currently supported for building hierarchies.
\bi
\item {\bf Bottom-up Intersection (BU-i)} \\
This method  is based on the intersection of the names of the original clusters.
At first we find the nodes  whose  names have the biggest (in size) intersection
and we group them by   creating a new node that has these nodes as children.
The name of the newly  created node   is the intersection of names of its children.
We continue analogously  until reaching a single node
(and in each step we ignore the nodes that have already fathers).

\item {\bf Bottom-up Weighted  (BU-w)} \\
    This method takes  into account the weights of the centroid vectors.
    At first we order the words of each individual cluster name based on their
   significance in decreasing order
   (in case of ties we order the corresponding  words alphabetically).
   Subsequently we order the resulting cluster names alphabetically.
   In this way the clusters that have the same  most weighted term(s)
   will   be placed consequently in the resulting order.
   Additional layers of internal nodes
   can be created by  grouping the clusters
   based on their names. Specifically
   two or more clusters are grouped under the same node
   if their names have a  common prefix (i.e. at least one common word at their beginning).

\item {\bf Top-Down (TD) } \\
    This is actually a divisive clustering method.
    The original $K$ clusters are considered as children of the root node.
    We continue clustering the contents of each cluster
    by reapplying K-means. We stop clustering a cluster
    when the maximum depth ($d_{mx}$) has been reached
    or when the size of that cluster is too small
    to be further partitioned   (less than  a threshold $sz_{mn}$).
\ei
Finding the configurations that work well in practice is subject for further research.
As one could expect,
if stemming is activated
then the cluster names
contain stems (instead of real words)
and this reduces their  quality (readability).

\comment{==========
    In brief,  the clusterer has  the following configuration parameters:\\
    $L$: number of documents to be clustered starting from the most highly ranked
        documents (default = 100), \\
    $K$: number of clusters (notice that in TD, $L$ should be much greater than $K$ if we want several levels),\\
    $tl_{mn}$ (resp. $tl_{mx}$):
        the minimum (resp. maximum) number of words that a cluster name could have
        (default value is 1 and 5 respectively),
        but the $w$ (for each given query) also affects these parameters
        (e.g. if $w=4$ and $t_{mn}$ has been set to 1, the real value used for $t_{mn}$ is 4)
        \\
    $sz_{mn}$:  the minimum number of docs in a cluster (default = 5), mainly useful for TD \\
    $d_{mx}$:  the maximum depth of the cluster tree (default =0), \\
    $w_{mx}$: the maximum number of words  in each document (most weighted words to be considered, default =20),
    and this affects the efficiency \\

    Table \ref{tbl:ClusteringParams} describes the applicability of the
    configuration parameters to the various clustering approaches

    \puttableSV{tbl:ClusteringParams}{Clustering Configurations}{}{
    {\small
    \btbl{|l|l|l|l|l|}\hline
    Parameter   & BU-i & BU-w & TD & Comment \\\hline
    $L$         & $\checkmark$ & $\checkmark$ & $\checkmark$ & \\\hline
    $K$         & $\checkmark$ & $\checkmark$ & $\checkmark$ &  in TD it should be $L >> K$\\\hline
    $tl_{mn}$   &            &            &            &  however if $w > tl_{mn}$ then  $tl_{mn}' = w$\\\hline
    $tl_{mx}$   &            &            &            &  however if $w > tl_{mx}$ then  $tl_{mx}' = w$\\\hline
    $sz_{mn}$   & $\times$     & $\times$     & $\checkmark$ & \\\hline
    $d_{mx}$    & $\checkmark$ & $\checkmark$ & $\checkmark$ & \\\hline
    $w_{mx}$    & $\checkmark$ & $\checkmark$ & $\checkmark$ & it affects the efficiency\\\hline
    \etbl
    }
    }

Table \ref{tbl:Clustering} gives some empirically satisfactory configurations
that we have identified.

\puttableSV{tbl:Clustering}{Clustering Configurations}{}{
\btbl{|l|l|l|l|l|l|l|l|l|}\hline
Method  & $L$  & $K$  & $tl_{mn}$ & $tl_{mx}$ & $sz_{mn}$ & $d_{mx}$ & $w_{mx}$ & example queries\\\hline\hline
BU-w    & 100  &  10  &     0     &     0     &     X     &    0     &    10    & \\\hline
BU-i    & 100  &  10  &     0     &     0     &     X     &    0     &    10    & \\\hline
BU-w    & 100  &  10  &     0     &     0     &     X     &    2     &    10    & this seems to work better\\\hline
BU-w    & 100  &  10  &     1     &     5     &     X     &    2     &    20    & \\\hline
BU-w    & 100  &  10  &     0     &     0     &     X     &    2     &    15    & \\\hline
TD      & 100  &  10  &     0     &     5     &     5     &    1     &    10    & \\\hline
\etbl
}

Indicative execution times: ...

======================}

\subsection{Automatic Taxonomy Construction}
\label{sec:Taxonomy}

\groogle\ can also organize the words that appear in its
lexicon in the form of a taxonomy.
A quite efficient method was designed for this purpose.
It consists of the following steps:
\bi
\item[(1)] Compute the minimum and maximum frequency of the words in the lexicon
   (denoted by $df_{mn}$ and $df_{mx}$ respectively).
\item[(2)] Partition the interval $[df_{mn},df_{mx}]$  into $L$  successive intervals
    (where $L$ is administrator-provided),
   i.e. $[df_{mn},df_{1}], \ldots, [df_{L-1},df_{mx}]$.
   We will refer to them with $lev_1, \ldots,  lev_L$ respectively.
\item[(3)] Ignore the intervals corresponding to  low frequencies,
   specifically keep only the $M$ intervals with the highest frequencies
   ($M$ is administrator-provided and it should be  $M<L$),
   i.e. keep only $lev_{L-M-1}, \ldots, lev_L$.
\item[(4)] Assign to  each of these $M$ intervals
    those words whose frequency falls to that interval.
\item[(5)] For each word $w_i$ of level $z$ (where $z \leq L-1$)
    connect it with  the most "correlated"  word of the level $z+1$
    (that word will be the "parent" of $w_i$).
    In this way we get a tree-structured hierarchy of words.
\ei
Regarding step (5),
the correlation $c_{ij}$ between
two words $w_i$ and $w_j$ is  computed using the formula:
\begin{equation}
c_{ij} = \sum_{d_k\in D} tf_{ik} \times tf_{jk}
\label{LABELLING}
\end{equation}
where
$tf_{ik}$ is the frequency of term $i$ in document $k$.

As an example,
Table \ref{tbl:Taxonomy}
describes
the partitioning
obtained assuming $L=20$
(for each level the table shows the number of words
that belong to that level).
To construct
the taxonomy
we have considered only the last 5 groups
(empty groups, like level 19, are considered as non existant).
So the taxonomy includes 35 words in total.
After creating the connections between words
we realized
that each word has an  average of 1.4 child nodes.

The reason for partitioning words into groups
(according to their frequency) is for avoiding
computing the correlation matrix between all
pairs of words
(which would be formidably expensive\footnote{
 That would require $N |V|^2/2 $ calculations
 where $N$ is the number of documents,
 and $|V|$ the size of the lexicon.
}).
In addition, ignoring those words that occur rarely
further improves efficiency
(as more than 95\% of the vocabulary has a very small document frequency)
and does not harm the quality of the result
	as
these words do not describe the main concepts of the document corpus,
and we have not anyway adequate statistical information to connect them right in a hierarchy.

\puttableSV{tbl:Taxonomy}{Partitioning the lexicon into two groups }{}{
\btbl{|l|l|l|l|l|l|l|l|l|l|l|l|l|l|l|l|l|l|l|l|l|}\hline
       & \multicolumn{20}{|l|}{Low frequency \hspace*{5cm} High frequency} \\\hline
Level  &  1 & 2 & 3 & 4 & 5 & 6 & 7 & 8 & 9 & 10 & 11 & 12 & 13 & 14 & 15 & 16 & 17 & 18 & 19 & 20  \\\hline
Num. Of Words    & 217142 & 1103 & 523 & 292 & 199 & 128 & 83 & 83 & 53 & 52 & 25 & 18 & 18 & 14 & 14 & 7 & 8 & 2 & 0 & 4  \\\hline
\etbl
}

The taxonomy is created after the end of crawling or on demand and
requires around 68 minutes.
The reason for this slow performance is that a vast number of queries is issued
to the database,
asking the $tfs$ of the terms in the top five levels for each document
(testing collection comprises of 14246 documents).

\bc
\ec

\subsection{Presentation of Results}
\label{sec:UI}

The final step of the retrieval process is the presentation of the results.
Contrary to popular web search engines, \groogle\
computes all the results at once.
For each page in the results,
a small surrogate is presented,
including the title of the page
and a short excerpt that we call {\em best text}.
This excerpt should ideally contain all words of the query.
To find such query-dependent excerpts
\groogle\ keeps a copy of the full text of the pages
(in addition to the index) at a cost of extra storage space.
For this purpose, after the lexical analysis of each document,
its full text is stored in a separate file in the local file system.
This yields a total overhead of about 9.77\% of the original collection size.

The query-dependent excerpt is computed as follows:
Since by default, we do not store term positions in the
index, we split the full text of relevant documents, shown in
the results page, into 10-grams of words. The final excerpt is
the concatenation of the two 10-grams that contain the most
occurences of query terms.

Finally, the user interface provides necessary components
(check boxes and radio buttons),
to support edit distance
and query expansion functionalities, previously
described. Additionally, in case the clustering option
has been selected by the user, the clusters are represented in
an accessible way by splitting the page in two frames. The left
frame contains the names of the clusters in tree-order while the
contents of the selected cluster are displayed on the right.
An indicative screen dump is shown in Figure \ref{fig:screenshot1}.

\putfigureSV{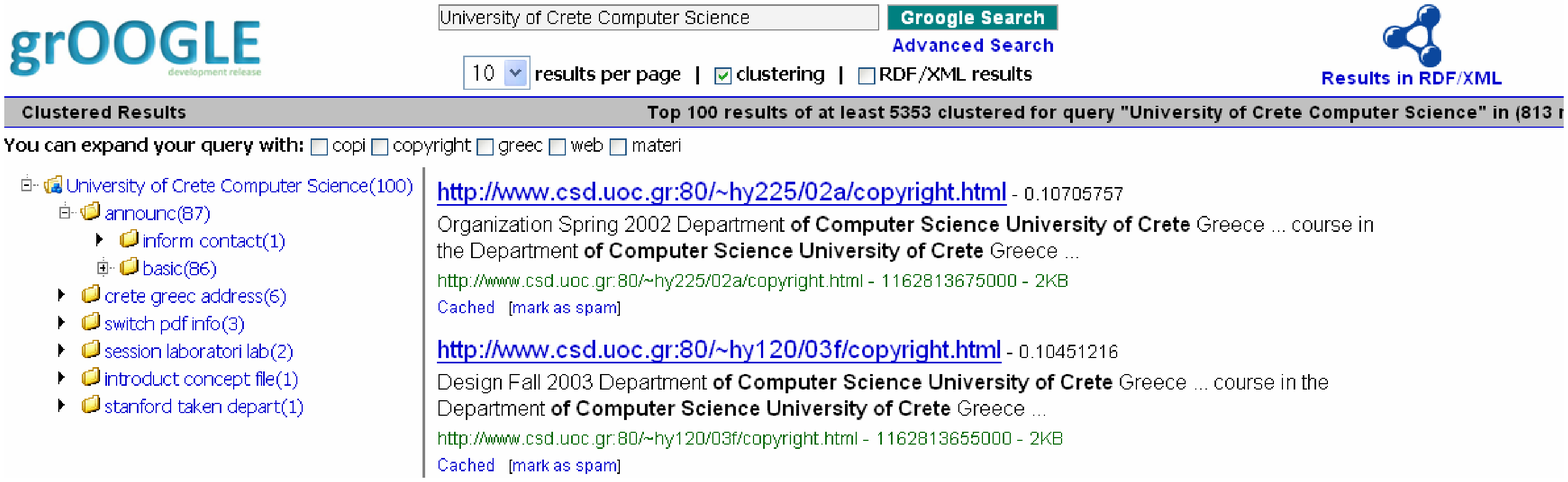}{37mm}{fig:screenshot1}
  {Results page}{}


\puttableSV{tbl:admin}{Administration functionalities}{}{
{\scriptsize
\btbl{|l|l|}\hline
$Modules$   &   $Functionalities$               \\\hline
Crawler     & Starting points (seeds)           \\
        & Collection name               \\
        & Accept/Reject filetype list           \\
        & Maximum downloaded pages          \\
        & Host/Domain spanning              \\
        & Traversal algorithm               \\
        & Maximum crawling depth            \\
        & Repository path               \\
        & Configuration file path           \\
        & Logging (file path, log level)        \\
        & Re-crawling time period           \\\hline
Indexer     & Create index                  \\
        & Drop index                    \\
        & Create new collection             \\\hline
Clustering  & Algorithm used                \\
        & Number of clusters                \\
        & Maximum number of documents to be clustered   \\
        & Max/min title length for clusters     \\
        & Maximum depth for cluster tree        \\
        & Name hierarchy for cluster tree       \\
        & Minimum documents in clusters         \\
        & Maximum number of words in each document  \\\hline
Taxonomy    & Number of levels              \\
        & Number of output levels           \\\hline
Query Expansion & Enable/Disable                \\
        & Number of relevant documents          \\
        & Number of relevant words          \\\hline
Query Evaluator & Edit distance tolerance           \\\hline
\etbl
}
}

\subsection{Administration}
\label{sec:Admin}

This module glues together the various components
and offers to the administrator of the web search
engine a Web-based user interface for controlling its
behavior, quickly and easily. Supported
functionalities are listed in Table \ref{tbl:admin}.

\section{Comparison with Other Systems}
\label{sec:RW}

\groogle\ offers a really wide spectrum of functionalities.
Just indicatively,
Table \ref{tbl:Comparison} lists the features
that \groogle\ and some well known IR systems and web search engines offer.


\puttableSV{tbl:Comparison}{System \& Functionalities Matrix}{}{
{\footnotesize
\btbl{|l|c|c|c|c|c|c|c|}\hline
Functionality       & \groogle\  & google      & Terrier v.1.1 & Vivisimo   & Yahoo!     & Lucene     & Lemur \\\hline\hline
Crawler             & $\checkmark$ & $\checkmark$  & $\checkmark$    & $\checkmark$ & $\checkmark$ & $\times$     & $\times$ \\\hline
Greek Stemmer       & $\checkmark$ & $\times$      & $\times$        & $\times$     & $\times$     & $\times$     & $\times$ \\\hline
Result Clustering   & $\checkmark$ & $\times$      & $\times$        & $\checkmark$ & $\times$     & $\times$     & $\checkmark$ \\\hline
Edit Distance       & $\checkmark$ & $\checkmark$  & $\times$        & $\times$     & $\checkmark$ & $\times$     & $\times$ \\\hline
Query Expansion     & $\checkmark$ & $\times$      & $\checkmark$    & $\times$     & $\checkmark$ & $\checkmark$ & $\checkmark$ \\\hline
Link Analysis       & $\checkmark$ & $\checkmark$  & $\times$        & $\times$     & $\checkmark$ & $\times$     & $\checkmark$ \\\hline
Several Retrieval Models & $\checkmark$ & $\times$ & $\checkmark$    & $\times$     & $\times$     & $\times$     & $\checkmark$ \\\hline
\etbl
}
}

\subsection{Indexing Performance}
\label{sec:Indexing}

We compared the indexing mechanisms of \groogle\ and Terrier
version 1.1 using
our collection (the one described in Section \ref{sec:Components}).
The entire collection is approximately 2.8 GB.
Block addressing and debug message printing were disabled in both tools.
Stemming and stop-words removal were also disabled in \groogle.
The results of the comparison are shown in Table \ref{tbl:ToolsComparison}.
Notice that the number of documents indexed by the two tools is not the same.
\groogle\ managed to index more documents than Terrier due to its better
file type identification technique.
 The files that \groogle\ failed to index are executables and media files,
constituting only the 3.8\% of the collection.

Regarding index size,
the index of Terrier is about 8 times smaller than that of \groogle.
Recall that \groogle\ uses a DBMS,
while Terrier uses compression
and special  data structures.
Nevertheless,
the index size of \groogle\
is less than the 10\% of the collection's size. The size may increase in the
future, if we decide to use the SP-GiST trie, since the B+-trees scale better
with respect to index size according to \cite{Eltabakh06}.

On the other hand,
\groogle\ offers faster index creation:
it takes less than 1 hour to index about 2.8 GB of
documents and is about four times faster than Terrier. In the future we will try 
to reproduce our findings with an updated Terrier version.

\puttableSV{tbl:ToolsComparison}{Comparison of Indexing Techniques}{}{
\btbl{|c|c|c|c|}\hline
$Tool$  & $Index\ Size$    & $Documents\ Indexed$    &   $Time$ \\\hline\hline
Terrier v.1.1   & 23.8 MB (0.8\%) & 11699 (80.6\%) & 11694 sec \\\hline
\groogle & 214.99 MB (7.4\%)& 14246 (97.2\%)    & 2892 sec   \\\hline
\etbl
}

\subsection{Searching Performance}
\label{sec:Searching}

Table \ref{tbl:QueryEvaluationTimes} reports query evaluation times.
These query evaluation times were calculated using the previously described
collection of $N$=14246 documents with a vocabulary size of $|$V$|$=220518.

\puttableSV{tbl:QueryEvaluationTimes}{Query Evaluation Times}{}{
{\scriptsize
\btbl{|c|c|c|c|}\hline
\multicolumn{1}{|c|}{\textbf{Model}} & \multicolumn{1}{|c|}{\textbf{q}} & \multicolumn{1}{|c|}{\textbf{$|$ans(q)$|$}}  & \multicolumn{1}{|c|}{\textbf{Time}}  \\\hline
\multicolumn{4}{|c|}{}\\\hline
Vector & csd math  & 1449 & 79 ms        \\\hline
Boolean & csd or math & 1449 & 58 ms         \\\hline
Extended Boolean & csd or math & 1449 &  187ms          \\\hline
Fuzzy & csd or math & 1449 &  165 ms           \\\hline
\multicolumn{4}{|c|}{}\\\hline
Vector &information retrieval systems  & 6329 & 315 ms        \\\hline
Boolean & information or retrieval or systems  & 6329 & 260 ms        \\\hline
Extended Boolean &information or retrieval or systems  & 6329 & 522 ms        \\\hline
Fuzzy & information or retrieval or systems  & 6329 & 557 ms        \\\hline

\etbl
}
}

Table \ref{tbl:QueryEvaluationComparisons} compares the evaluation times and results of \groogle\ and Terrier,
using the Vector Space model and the collection previously described. 100 random queries were executed on both engines and
average times were calculated. The process was repeated for 1 up to 10 random query terms. The modulus of the average difference
in the number of the results was also calculated for each case.

Terrier offers constant searching times, regardless of terms number in the query
and much higher performance than \groogle\ (due to the not optimized and DBMS nature of \groogle). On the other hand \groogle\ times
increase monotonically with the number of terms in the query. Note the difference in the results number, a consequence of 
the bigger number of documents indexed by \groogle.

\puttableSV{tbl:QueryEvaluationComparisons}{Average Evaluation Times and Result Differences for 100 Iterations}{}{
\btbl{|c|c|c|c|}\hline
$Query\ Terms\ Num.$	&	$\groogle\ Avg.\ Time$	&	$Terrier\ Avg.\ Time$ & $|Result\ Avg.\ Diff.|$ \\\hline\hline
1	&	0.0690	& 	0.0018	&	2.35	\\\hline	
2	&	0.1598	&	0.0020	&	10.01	\\\hline
3	&	0.1540	&	0.0022	&	14.94	\\\hline
4	&	0.1741	&	0.0014	&	7.12	\\\hline
5	&	0.2016	&	0.0016	&	28.79	\\\hline
6	&	0.2005	&	0.0018	&	52.4	\\\hline
7	&	0.2434	&	0.0018	&	78.24	\\\hline
8	&	0.2714	&	0.0014	&	5.16	\\\hline
9	&	0.3284	&	0.0017	&	56.71	\\\hline
10	&	0.3212	&	0.0044	&	90.23	\\\hline
\etbl
}

\section{Summary and Concluding Remarks}
\label{sec:Concl}

There are only few concise papers (e.g. \cite{arasu2001sw})
that discuss all aspects revolving the design, implementation and evaluation
of Web search engines.
This paper presented a brief description of \groogle,
a currently developed web search engine
offering a wide spectrum of functionalities
including
 an advanced stemmer for the Greek language
 and real-time result clustering.
Another distinctive characteristic of \groogle\
is that its index is based on a DBMS.
For this reason,
we have discussed issues regarding the usage of DBMSs
for managing the index
and  we have compared our engine with other engines
whose index is not based on a DBMS.
In brief, the DBMS-based index
requires  more
storage space and offers poorer performance (as expected),
but on the other hand it is more extensible.
In addition:
\bi
\item we realized that stemming has a number of benefits (improves recall, reduces the size of the index),
  but affects negatively the selection  of the best text (that is used for creating  document surrogates),
  the quality (readability) of the cluster names,
  and the quality of the suggested terms (during query expansion),
\item we verified that the distribution of words approximates a power-law distribution,
\item we realized that result clustering is a hard problem that is worth further research.
\ei
In future, we plan to
\bi
\item conduct experiments with larger collections.
\item extend the crawler so that to support important-first-traversal
  (also taking into account the rank field in the db).
\item continue improving the real-time results clustering module
\item investigate  PostgreSQL indices that
     offer better performance for indexing and query evaluation.
\item reproduce indexing times with an updated version of Terrier and a bigger documents collection.
\ei

\section*{Acknowledgements}
\label{sec:CR}

The engine was developed as a student project in the IR course (CS463)
at the Computer Science Department of the University of Crete
in two semesters (spring 2006 and spring 2007).
Many thanks to all students that have contributed:
Evangelos Boutsakis, Nikos Dimaresis, Stefanos Dubulakis, Dimitra Emmanouilidou, Manos Frantzolakis,
Giorgos Georgopoulos, Katerina Gkirtzou, Nikos Grispos, Nikos Kampitakis, Kostas Kapakiotis,
Stelios Kapetanakis, Giorgos Konstantinidis, Manos Kritsotakis, Michael Markogiannakis,
Antonis Melakis, Yiannis Papadakis, Kostas Perakakis, Kyriakos Sidhropoulos, Apostolos Stamou, Manos Tavlas
and Axilleas Tziatzios.

\bibliographystyle{plain}
\bibliography{paper}

\end{document}